\documentstyle [fleqn,11pt] {article}
\newcommand{\cd}{\cdot}

\newcommand{\al}{\alpha}
\newcommand{\de}{\delta}
\newcommand{\De}{\Delta}
\newcommand{\ep}{\epsilon}
\newcommand{\ga}{\gamma}
\newcommand{\Ga}{\Gamma}

\newcommand{\La}{\Lambda}

\newcommand{\Om}{\Omega}
\newcommand{\om}{\omega}
\newcommand{\si}{\sigma}
\newcommand{\Si}{\Sigma}
\newcommand{\th}{\theta}

\newcommand{\mm}{\mbox{$\cal M$}}

\newcommand{\lap}{\triangle}

\newcommand{\bm}[1]{\mbox{\boldmath $#1$}}

\newcommand{\be}{\begin{equation}}
\newcommand{\ee}{\end{equation}}
\newcommand{\bea}{\begin{eqnarray}}
\newcommand{\eea}{\end{eqnarray}}
\newcommand{\bean}{\begin{eqnarray*}}
\newcommand{\eean}{\end{eqnarray*}}
\newcommand{\dd}{\partial}

\begin{document}

\hfill  ZU--TH  28/93
\begin{center}
{\LARGE {\bf Early Reionization in Cosmology}}\vspace{1.5mm}\\
	{\large R. Durrer \\
	Universit\"at Z\"urich\\
	Institut f\"ur Theoretische Physik \\
        Winterthurerstrasse 190\\
	CH--8057 Z\"urich\\
	SWITZERLAND}
\end{center}

\section{Introduction}

Since this is the first contribution to this meeting which is mainly
involved in cosmology,
I would like to put it into its cosmological perspective:

In the standard model of cosmology we assume that the universe is
homogeneous and isotropic on large enough scales. This assumption,
originally probably made just by reasons of simplicity, leads to a so
called Friedmann Lema\^{\i}tre universe which explains, e.g., the well
established uniform Hubble expansion \cite{Sa}. Within this model, the
universe started out from an extremely dense, hot initial state and
subsequently cooled  by adiabatic expansion undergoing a series of
phase transitions. At a temperature of about 0.1MeV$\approx 10^9$K,
deuterons become stable and virtually all the neutrons present are bound
into $^4$He. This leads to the well established abundances of the
light elements. As the universe cools further, below about $3000^o$K
there are no longer enough ionizing photons around to keep the Hydrogen
Helium plasma ionized. The matter in the universe recombines to neutral
H/He and the universe becomes transparent to the cosmic radiation field.
Radiation can then propagate freely, influenced only by the cosmic
expansion and redshifted to the $2.7^o$K microwave background which we
observe today. At recombination, the age of the universe was
$t\approx 2\times 10^5$years,
which corresponds to a redshift of $z_R\approx 1100$.

We now want to discuss the possibility, that the cosmic plasma
might have reionized again at some lower redshift $z_i<z_R$.
The reason one might want to investigate this idea is twofold:
\begin{itemize}
\item {\bf Observationally:} The Gunn Peterson  test, i.e. the absence
 of a Lyman-$\al$ trough in all the observed quasar spectra, shows that
 the intergalactic medium is ionized for $z< 4.5$ \cite{GP,FB}
\item {\bf Theoretically:}  The anisotropies in the cosmic microwave
 background (CMB)  have turned out to
 represent one of the most stringent 'bottle necks' which a scenario of
 structure formation has to pass in order to be acceptable. As a
 possibility  to relax this constraint, it has been proposed
 that early reionization can damp CMB  fluctuations on small scales
 ($\th\le 6^o$) due to photon diffusion in the ionized
 plasma \cite{Pe}. We shall illustrate this idea in this paper.
\end{itemize}

In the next section, we want to explain the influences of reionization on the
cosmic microwave background: It leads to damping of fluctuations due to
photon diffusion on one hand and slightly distorts the CMB spectrum on
the other hand. In  Section~3, we  discuss  the
processes in the plasma  which have to be taken into account to
describe  reionization and formulate the system of differential equations.
Finally we draw conclusions and outline future progress.

In this contribution we use the following notation:
\begin{itemize}
\item Greek indices run from 0 to 3 and latin ones from 1 to 3
\item We assume throughout a spatially flat universe, i.e. the density
   parameter, $\Om_{tot}=\rho_{tot}/\rho_c=1$
   (a possible curvature would only change {\em angles}, but no other
   conclusions since we are mainly interested in the regime $z>>1$).
\item We choose $t$ as conformal time coordinate and work with the metric
  signature $(-,+,+,+)$, so that $ds^2 = a(t)^2(-dt^2+d\bm{x}^2)$.
\item Boldface characters denote 3 dimensional vectors.
\item We parametrize  Hubble's constant by $H_0 = 100$km/(s Mpc)$h$.
  The density parameter of the baryons, is denoted by
	$\Om_B = \rho_B/\rho_c =
	\rho_B/(7.7h_{50}^2\times 10^{-29}g/cm^3)$.
 Observations limit $0.5<h<0.8$ and (including nucleosythesis calculations)
 $0.01<\Om_B<0.1$.
\end{itemize}

\section{Reionization and the cosmic microwave background}

\subsection{Damping of fluctuations by photon diffusion}

As we shall see in this paragraph, reionization can lead to a
substantial damping of fluctuations in the CMB on small angular scales.
This is important for some scenarios of structure formation to overcome
present limits posed by small and medium angular scale experiments (see
contribution of P. Richards in this proceedings).
As an example, we mention the recently investigated scenario with
cold dark matter (CDM) and texture seeds \cite{Tu,TS,d90,STPR,d93}.
There, an analysis of
CMB anisotropies shows that early reionization is a crucial ingredient
for this scenario \cite{DHZ}. Without damping, the
small scale anisotropies would dominate and  exceed observed limits. On
the other
hand, studies of the texture scenario show, that textures lead to early
formation of objects \cite{d90}. At a redshift $z \approx 50$ about $1\%$
of the baryons in the universe have collapsed and formed objects with
mass $M_{nl}\le 10^5M_\odot$ \cite{GST}. If one assumes that,
due to the formation of these objects, radiation energy of about 100keV per
nucleon is emitted, this would yield a total  energy density of this
ionizing radiation $\rho_{i}=
 q\rho_B \approx 10^{-6}\rho_B \approx 10^3eV\times n_B$, which is by
far enough to reionize all the hydrogen in the universe.

By early reionization in this context we mean that it has to happen early
enough so that Compton scattering of electrons is still effective. At late
times electrons are too sparse to scatter photons effectively and the
electron proton plasma is again decoupled from radiation.  This decoupling
time is determined by the optical depth due to Compton scattering being
equal to unity:
\[   \tau = \int_{t_0}^{t_{dec}}n_e\si_Tadt  = 1 ~,  \]
which leads to a decoupling redshift of
\be   z_{dec} = 100({0.025\over \Om_Bh})^{2/3} ~, \label{1zd}
\ee
where $\si_T$ denotes the Thomson
cross section. We further assume (for simplicity) $\Om_{tot}=1$, so that
a substantial amount (about 90-95\%) of non baryonic, dark matter has to
be present. We parametrize the metric of this spatially flat Friedmann
universe by using conformal time,
\be ds^2 = a^2(-dt^2+d\bm{x}^2) ~~.\ee
The physical time differential is thus given by $adt$.

We now want to investigate how CMB anisotropies on scales smaller than
$t_{dec}$, i.e., $ \th < 1/\sqrt{z_{dec}+1} \approx 6^o$ can be damped
if reionization happens substantially before $z_{dec}$.

Boltzmann's equation for Compton scattering  is given by
\be p^{\mu}\dd_{\mu}f-\Ga^i_{\mu\nu}p^\mu p^\nu{\dd f \over \dd p^i} =
  C[f] ~ , \label{2boltz} \ee
where $f$ denotes the distribution function of the photons and $C[f]$ is
the collision integral.
In a perturbed Friedmann universe we separate $f$ into an isotropic
background contribution and a small perturbation $f= \bar{f}+\de f$.
One can then find a gauge--invariant (i.e. invariant
under linearized coordinate transformations) variable  $\cal F$, which
reduces to $\de f$ for perturbations which are much smaller than the size
 of the horizon. For the energy integrated ``brightness perturbation''
\[ {\cal M} = {4\pi\over \rho}\int_0^\infty {\cal F}p^3dp \]
the perturbation of the Boltzmann equation (\ref{2boltz}) then becomes
(for scalar perturbations)
\be \dot{\cal M} +\ep^i\dd_i\mm = 4\ep^i\dd_i(\Phi-\Psi) +
    a\si_Tn_e[D_g^{(r)}-\mm -4\ep^il\dd_iV + {1\over 2}\ep_{ij}M^{ij}]
         \; . \label{2Bb} \ee
Here $\bm{\ep}$ is a unit vector which denotes the direction of the
photon momentum, $V$
is a potential for the baryon velocity,  $\Phi$, $\Psi$ are the so called
Bardeen potentials which parametrize scalar perturbations of the geometry
and $l$ is an arbitrary length scale introduced merely to keep the perturbation
variables dimensionless.
Furthermore
\bean  \ep_{ij} &=& \ep_i\ep_j - {1\over 3}\de_{ij} ~,\\
   D_g^{(r)} &=& (1/4\pi)\int{\cal M}(\ep)d\Om  ~~~\mbox{ and }~~  \\
   M^{ij} &=&  {3\over 8\pi}\int\mm(\ep)\ep_{ij}d\Om]~. \eean
The baryon equation of motion is
\be l\dd_j\dot{V} + (\dot{a}/a)l\dd_iV = \dd_i\Psi -
        {a\si_Tn_e\rho_r\over 3\rho_b}(M_j +4l\dd_iV) ~, \label{2bar}\ee

\[  \mbox{ with }~~~ M_j ={3\over 4\pi}\int\ep_j\mm d\Om ~.\]

An introduction to gauge--invariant cosmological perturbation theory and a
thorough derivation of eqs. (\ref{2Bb},\ref{2bar}) is given in \cite{d93}.

In eq. (\ref{2Bb}) the first contribution to the right hand side is the
gravitational force. In the second contribution, the collision integral,
 the first two terms are the usual smoothing, then there is a Doppler term
and the final term is due to the
anisotropy of the Compton cross section (it can be neglected in the limit of
very many collisions). In eq. (\ref{2bar}) we see, that on the right hand side
the drag force due to the radiation drag experienced by the electrons is added
to the well known gravitational acceleration term.

We now want to estimate the damping of $\cal M$ due to photon diffusion.
Setting $t_T = (a\si_Tn_e)^{-1}$, we investigate the limit $t_T/t << 1$.
In deepest order we find
\[ \dot{D}_g^{(r)} = (4/3)l\lap V = (4/3)\dot{D}_g^{(B)} ~,\]
where $D_g^{(B)}$ is a perturbation variable for the baryon energy density
perturbation, and the second equal sign is due to the baryon
number conservation equation. In this limit therefore, the entropy per
baryon is conserved, $\de s/s=\de n_B/n_B$, i.e., baryons and radiation
behave like a single perfect fluid with energy density $\rho_r+\rho_B$
and pressure $p_r=(1/3)\rho_r$. We now want to analyse the system
(\ref{2Bb},\ref{2bar}) in second order. Let us first neglect the time
dependence of the coefficients and make a plane wave ansatz
\[ V \propto {\cal M} \propto e^{i(\bm{k\cd x}-\om t)} ~.\]
Inserting this ansatz into (\ref{2Bb},\ref{2bar}), we obtain the
following dispersion relation in lowest order $kt_T$ and $\om t_T$, i.e.,
for short waves;
\be \om = \om_0 +i\ga ~~,~~\om_0={k\over \sqrt{3(1+R)}} ~~,~~~
    \ga = {k^2t_T\over 6}{(R^2 + (4/5)(R+1))\over (R+1)^2}~, \label{2disp}\ee
with $R = 3\rho_B/4\rho_r$ \cite{Pe,d93}. In the matter
dominated regime, $R>>1$,
we have $\ga \approx  {k^2t_T\over 6}\approx {2\pi t_T\over l^2}$ where
$l$ denotes the wavelength of the perturbation.

We then approximate the total damping by
\[ e^{-f} ~~\mbox{ with }~~~~ f = \int_{t_{in}}^{t_{end}} \ga(t)dt ~,\]
where $t_{end}$ is defined by $t_T(t_{end})= l/2\pi$ or $t_{end}=t_{dec}$
(whatever condition is satisfied earlier); and $t_{in}$ is the time when
the perturbation enters the horizon $t_{in} \approx l/2$ or the time of
reionization, $t_{in}=t_i$ (whatever happens later).

If we use this estimates for damping of fluctuations induced by a spherically
symmetric collapsing texture \cite{Tu,d90,d93}, we obtain
\be f = \left\{    \begin{array}{ll}
\left({1+z_{dec}\over 1+z_c}\right)^{3/2}\left[\left(
      {1+z_{dec}\over 1+z_c}\right)^{15/8} -1\right] & \mbox{ for }
	z_c>z_{dec}\\
      0  & \mbox{ else,} \end{array} \right. \label{2f}\ee
where $z_c$ denotes the redshift of texture collapse,
$t(z_c)\equiv t_c \approx l$ and we have assumed $z_i>z_c$.
This naive estimate, which leads to a factor $\approx 15$  damping for
$z_c =90z_{dec}~~ (t_c = 0.1t_{dec})$;  a factor $\approx 5$ damping for
$z_c =10z_{dec}~~ (t_c = 0.3t_{dec})$ and a factor $\approx 1.6$ of
damping for $z_c =2z_{dec}~~ (t_c = (1/\sqrt{2})t_{dec})$.

In Fig.~1-4 we show the results from numerical integration of the system
(\ref{2Bb},\ref{2bar}), with a gravitational field induced from a
spherically symmetric,  collapsing texture \cite{DHZ}.
Comparisons with the naive predictions (\ref{2f}) show that our
approximation  is actually quite reasonable!

This result tells us also that we must require at least $z_i> 2z_{dec}$ to
obtain damping by, say, a factor of 2.

\subsection{Spectral distortion}
In addition to damping the amplitude of fluctuations, the reionized plasma
also distorts CMB spectrum: The ionizing radiation does not only ionize matter
but also heats up the plasma to temperatures around typically $T_e\approx$
0.3 to a few eV.  Nonrelativistic Compton scattering up scatters the low
energy microwave photons. If the Plasma has a thermal (Boltzmann) distribution
and the Photons are Planck distributed (as in our situation), the induced
change in the spectrum can be described by a single parameter, the
Compton--y parameter \cite{SZ7,SZ8}:
\be y \equiv {\si_T\over m_e}\int_{t_i}^{t_0} n_e(T_e-T_{CMB})a dt \approx
      0.4\times 10^{-5}\left({1+z_i\over 1+z_{dec}}\right)^{3/2}
	(\overline{T}_e/2eV)  ~,\label{2y} \ee
where $\overline{T}_e$ is a weighted  ``mean electron Temperature'':
\[ \overline{T}_e = (z_i+1)^{-3/2}\int_{0}^{z_i}T_e (z+1)^{1/2}dz ~. \]
(This approximation is excellent for $T_e>>T_{CMB}$ and $z_i>>1$, i.e. the
situation we are interested in.)

The observational limit set by the FIRAS experiment on COBE (see J. Mather,
this Proceedings) is
\be y < 2.5\time 10^{-5}  ~. \label{2yobs} \ee
In the next section we shall see, that ionization can be maintained by
collisions only if $T_e\ge 1.5$eV and, on the other hand, it seems to
require quite some fine tuning to maintain the plasma ionized by
photoionization without heating it  up to a temperature of about
(1 -- 2)eV. If we therefore assume a mean electron temperature of
\[\overline{T}_e \ge 1.5eV~, \] observation (\ref{2yobs}) already limits
the ionizing redshift to
\[ z_i \le 4z_{dec}~.\]

Improving the observational limit of the y--parameter by about a factor of 5
would therefore rule out reionization which happens early enough to lead
to significant damping of CMB fluctuations ($z_i>2z_{dec}$, say)!

\section{Requirements for an ionizing radiation field}
We now assume that at some high redshift $z>>z_{dec}$ there exists an
ionizing radiation field which  originates from the first
 nonlinear density perturbations, i.e. the first
'macroscopic' baryonic objects in the universe formed by
some unknown mechanism, and which is capable of reionizing the universe.
We want to study the properties of this radiation.

Its energy density $\rho_i$ and its spectrum $f_i$ are related by
\be \rho_i = {1\over\pi^2}\int_0^\infty\om^3f_i(\om)d\om ~~~~
   (\hbar = c =k_B = 1) ~ .  \label{3rhoi} \ee
We set
\[ \rho_i = q \rho_B \]
and assume that $q$ is slowly varying over some time period,
where $\rho_B$ is the baryon energy density. For this radiation to be
able to ionize the universe, we must require

\be n_i(\om>\De) \ge n_B ~~~~~ (\De = 1Ry = \al^2m_e/2= 13.6eV), ~\mbox{ where}
   \label{3cond} \ee
\[ n_i = {1\over\pi^2}\int_0^\infty\om^2f_i(\om)d\om ~ , \]
$\al$ denotes the fine structure constant and $m_e$ is the electron mass.
To be specific, let us assume that $f_i$ is a Planck spectrum with
chemical potential $\mu>> 1$,
\[ f_i = {1\over e^{\om/T_i +\mu} -1} ~ .\]
For this spectrum one finds
\[ \rho_i \approx {6\over \pi^2}T_i^4 e^{-\mu} ~~\mbox{ and thus }~~~
   q = {6 T_i^4 e^{-\mu} \over \pi^2 \rho_B} ~. \]
The requirement (\ref{3cond}) then yields
\be q > 0.8\times 10^{-7}e^{(\De/T_i)}
	[ (\De/T_i)^3-2(\De/T_i)^2 +2(\De/T_i)]^{-1} \label{3con2} \ee
This condition is shown by the heavy line in Fig.~5. It is compared with
observational limits on background radiations at the wavelengths
corresponding to the temperatures given in Table~1. The lower light line,
which corresponds to those limits today, $z=0$, shows that if there
exists a unresolved ionizing radiation background today its energy
has to be close to about $2eV$ and its intensity is only little
(less than a factor of 3) below the observational limit. In the
upper light line we translate the limits of Table~1 to $z=10$ (multiply
by $z+1$). The dashed region is the allowed parameter range if the
universe is to be photoionized until $z=10$, i.e., until recombination
is slower than expansion and the universe remains ionized without a
ionizing radiation.

\begin{table}
\begin{tabular}{|l|l|l|l|r|} \hline
 $\om_0$[eV] & $\Om_0$ & $q/10^{-7}$ & $T_i/\De/$ & Refs. \\ \hline
$2\cd 10^4$ & $10^{-9}$ & $0.1(z+1)$ & $1.5\cd 10^3\cd (z+1)$ &
	Setti '90 \cite{Se} \\
$2\cd 10^3$ & $10^{-9}$ & $0.1(z+1)$ & $150\cd (z+1)$ & Setti '90
	\cite{Se}\\
$200$ & $2\cd 10^{-10}$ & $0.02(z+1)$ & $15\cd (z+1)$ & Paresce
	\& Stern '81 \cite{PS}\\
$100$ & $6\cd 10^{-10}$ & $0.06(z+1)$ & $ 7.6(z+1)$ & Paresce
	\& Stern '81 \cite{PS}\\
$20$ & $7\cd 10^{-10}$ & $0.07(z+1)$ & $1.5\cd (z+1)$ & Paresce
	\& Stern '81 \cite{PS}\\
$8$ & $4\cd 10^{-8}$ & $4(z+1)$ & $0.6\cd (z+1)$ & Holberg '86 \cite{Ho} \\
$2.8$ & $2\cd 10^{-7}$ & $20(z+1)$ & $0.2\cd (z+1)$ &
	Longair '90 \cite{Lo} \\
$0.9$ & $1.2\cd 10^{-6}$ & $120(z+1)$ & $0.07\cd (z+1)$ &
	Longair '90 \cite{Lo} \\
$0.5$ & $3\cd 10^{-6}$ & $300(z+1)$ & $0.04\cd (z+1)$ &
	Longair '90 \cite{Lo} \\
$0.3$ & $3\cd 10^{-6}$ & $30(z+1)$ & $0.02\cd (z+1)$ &
	Longair '90 \cite{Lo} \\
$0.2$ & $10^{-7}$ & $10(z+1)$ & $0.015\cd (z+1)$ &
	Longair '90 \cite{Lo} \\
\hline \end{tabular}
\caption{Limits on diffuse background radiation from X--rays down
 to the near infrared are listed with the corresponding references.
The precipitous rise of the limits below 10eV is due to contributions
from airglow and zodiacal light}
\end{table}

\subsection{Processes in the plasma}
We now calculate the ionization and recombination rates and the heating
and cooling functions for the cosmic medium, neglecting the
Helium contribution:\\
{\bf Photoionization:} The Karzas--Latter Photoionization cross section
is given by
\be \Si_{pi} = {64\pi \over m_e^2\al 3\sqrt{3}}(\De/\om)^3g_{bf} ~,\ee
where $g_{bf}$ is a Gaunt factor which we set equal to 1 in the sequel
\cite{RL}.
The photoionization rate is correspondingly

\[ t_{pi}^{-1} = {8\pi m_e\al^5\over3\sqrt{3}}(1-x)\int_\De^\infty
  \om^{-1}f_i{d\om\over\pi^2} ~,\]
where $x= n_e/n_B$ is the ionization fraction.
Inserting now the thermal spectrum defined before we obtain
\be t_{pi}^{-1} \approx 6s^{-1}\times q(1-x)(\De/ T_i)^4E_1(\De/T_i)
    ({\Om_Bh^2\over 0.025})({1+z\over 200})^3  ~. \label{3pi}\ee
Here $E_1$ denotes the usual integral exponential function
\[ E_1(x) = \int_x^\infty {e^{-x}\over x}dx ~ .\]

The photons do not only ionize the plasma, but with their remaining energy
they also heat it up. This photoionization heating, $\Ga_{pi}$ is given
by

\bea \Ga_{pi} &=& {8\pi m_e\al^5\over 3\sqrt{3}}(1-x)\int_\De^\infty
  {\om-\De\over \om}f_i{d\om\over\pi^2} ~, \nonumber\\
  &\approx& 78(eV/s)\times q(1-x)({\Om_Bh^2\over 0.025})({1+z\over 200})^3
     (\De/T_i)^3[e^{-\De/T_i}-(\De/T_i)E_1(\De/T_i)] ~. \label{3Gpi}\eea

{\bf Recombination:}  We obtain the
 the recombination cross section from a detailed balance argument and the
photoionization cross section \cite{Sp}. This leads to the
recombination rate

\bea t_r^{-1} &=& {\De^{1/2}2^{13/2}\pi^{1/2}\al\over m_e^{3/2}3^{3/2}}
 	\left({3T_e/2\De +1\over (T_e/\De)^{1/2}}\right)x^2n_B
\nonumber\\
 &\approx& 10^{-13}s^{-1}\times (\De/T_e)^{1/2}x^2
	 ({\Om_Bh^2\over 0.025})({1+z\over 200})^3  ~. \label{3r}\eea

Due to the loss of one free particle, recombination leads to a loss in
kinetic energy, a contribution to the cooling function $\La$:
\be \La_r = (3/2)T_et_r^{-1} = 2\times 10^{-12}{eV\over s} (T_e/\De)^{1/2}x^2
	 ({\Om_Bh^2\over 0.025})({1+z\over 200})^3  ~. \label{3Lr}\ee

{\bf Collision:} Collisions of energetic electrons can lead to ionization
and/or excitation of hydrogen. For high enough electron temperatures this
is a very efficient cooling mechanism even is only a small fraction of
neutral hydrogen is present. The collisional ionization rate is about
\cite{Lo,La}
\be t_c^{-1} = 6\times 10^{-8}s^{-1}\times (T_e/\De)^{1/2}e^{-\De/T_e}x(1-x)
	 ({\Om_Bh^2\over 0.025})({1+z\over 200})^3  ~. \label{3c}\ee

The collisional ionization cooling rate is
\be \La_c = 8\times 10^{-7}{eV\over s}\times (T_e/\De)^{1/2}
	e^{-\De/T_e}x(1-x)
	 ({\Om_Bh^2\over 0.025})({1+z\over 200})^3  ~. \label{3Lc}\ee
Correspondingly, for  the excitation cooling one obtains
\be \La_e = 10^{-6}{eV\over s}e^{-3\De/4T_e}x(1-x)
	 ({\Om_Bh^2\over 0.025})({1+z\over 200})^3  ~. \label{3Le}\ee

Excitation cooling is thus always nearly $10^3$ times faster than
ionization cooling. Hence, the latter can be neglected.

As long as the electron temperature is substantially below 1eV,
excitations
are not common enough to cool the plasma efficiently. In this situation
Compton cooling of the CMB photons is the most efficient cooling mechanism.

{\bf Compton cooling:} From the Kompaneets equation one obtains

\bea n_B\La_{CMB} &=& {T_e-T\over T}({n_e\si_T\over \pi^2m_e})\int_0^\infty
	\om^4f_{CMB}(f_{CMB}+1)d\om  \nonumber \\
  &\approx& 10^{-10}{eV\over s} x\left({T_e - T_{CBM}\over\De}\right)
	\left({1+z\over 200}\right)^4  ~. \label{3LCMB}\eea

The corresponding heating process by compton scattering of the
ionizing radiation can be neglected (since there are so much less
ionizing photons than CMB photons).

Of course the plasma is also adiabatically cooled due to expansion and it is
important to compare this with the cooling rates above. Furthermore, if
the process of expansion is much faster than one of the processes calculated
above, we can neglect the corresponding process in the expanding universe.

{\bf Expansion:} In a matter dominated Friedmann universe with
$\Om_{tot}=1$, the expansion rate is  given by
\be t_{exp}^{-1} = H \approx 4.6\times 10^{-15}s^{-1}\times
	(h/0.5)\left({1+z\over 200}\right)^{3/2} ~. \label{3exp} \ee
Adiabatic cooling due to expansion amounts to a cooling rate of
\be \La_{exp}  = (3/2)(1+x)\left({dT_e\over dt}\right)_{ad} \approx
	2\times 10^{-13}{eV\over s}\times (T_e/\De)(1+x)
	\left({1+z\over 200}\right)^{3/2}(h/0.5) ~. \label{3Lexp} \ee

An additional possible cooling mechanism would be Bremsstrahlungs cooling
but this turns out to be very small within the range of electron
temperatures we are interested in.

\subsection{The differential equations}
The degree of ionization, $x$, the electron temperature, $T_e$, and the
form of the ionizing spectrum, $f_i$ are in principle determined by
the following system of differential equations:

\bea {dx\over dt} &=& -t_r^{-1} +t_{pi}^{-1} + t_c^{-1} \label{3Dx} \\
  {dT_e\over dt} &=& 2{\dot a\over a}T_e +{2\over 3(1+x)}
   (\Ga-\La) -{T_e \over 1+x}{dx\over dt}  \label{3DT}\\
{\dd f_i\over\dd t}-{\dot{a}\over a}{\dd f_i\over \dd \om} &=&
 -{df_i\over dt}|_{pi}  +{df_i\over dt}|_{rec}
 -{df_i\over dt}|_{Kompaneets}  + source ~ .\label{3Df} \eea

The first term on the right hand side of the second equation is
adiabatic cooling due to expansion. The second term denotes the other
heating and cooling mechanisms:
\[ \Ga = \Ga_{pi}~,~~~ \La = \La_r+\La_e+\La_{CMB} ~. \]
The last term is due to the increase of independent particles by ionization
 which lowers the  kinetic energy per particle and therefore the
temperature. The third equation sketches the changes in the ionizing
spectrum which are relevant at energies $\om\ge\De$. The second term on the
left hand side accounts for the redshift of photon energies due to
expansion. The first term describes the loss of photons due to ionization.
Then there comes the gain of photons by recombination, the energy losses
due to Compton scattering off the lower temperature electrons and finally
the source term which represents the emission of ionizing radiation from
some unknown primordial objects. This term has to be guessed.

We have not yet managed to solve the system of differential equations
(\ref{3Dx},\ref{3DT},\ref{3Df}) in some generality. We have just looked at
the stable situation, ${dx\over dt}={dT_e\over dt}={df_i\over dt}=0$ and
solved the resulting algebraic equations for $x$ and $T_e$ for a given
 ionization spectrum $f_i$ which we parametrized like in Section~2
as a Planck spectrum with chemical potential. $f_i$ is then fully
determined by its temperature $T_i$ and its chemical potential $\mu$ or
the parameter $q$ which was introduced in Section~2. The resulting electron
temperature and the degree of ionization are shown as functions of $T_i$ in
Figs.~6 and 7.
Of course this is not a realistic situation since photoionization is
so fast that it will quickly lead to a depletion of the spectrum above
13.6eV which is probably not readily refilled by a realistic source term.
This leads to the very unphysical behavior that even for very small
$T_i$, i.e. very few ionizing photons, $x\approx 1$.

\section{Results, Conclusions}
Our preliminary results show that early reionization is still possible
even though the parameter space is  squeezed by the limit on the
Compton y parameter and the absence of any near infrared background.
The first fact tells us that the electron temperature cannot have been
higher than 2eV for a substantial duration at high redshift, $z>z_{dec}$.
The second constraint reveals that, on the other hand, there was no
ionizing uv--radiation present at $z\approx 10$ which might have kept the
intergalactic medium ionized. If is was ionized at this redshift it thus
was collisionally ionized, i.e., the electron temperature was above 1.5eV.

To weaken these constraints,
one might still imagine the cosmic plasma to be photoionized up to
a redshift $z<z_{dec}$, say $z\approx 50$ and collisionally ionized
later. But it is not clear if even this possibility remains, i.e. if
it is possible to constantly photoionize the plasma without heating it
above 1.5eV, say. To decide on this last possibility we have to fully
solve the system of differential equations presented in Section~3.2 for
some reasonable source models. In addition we have to take into account
the clumping which must be present at a time when the first formed
objects  are supposed to emit ionizing radiation: Whenever a factor
$n_B^2$ enters our equations (e.g. recombination and collisions) we
have to replace the background Friedmann value of $n_B^2$ by
\[ n_B^{(clump)} = n_B^2(1+ \int \de(x)^2 d^3x) =n_B^2(1+\int P(k) d^3k),\]
where $\de(x)$ denotes the density fluctuations in the baryon distribution
and $P(k)$ is the power spectrum, $P(k)= (\hat{\de}(k))^2$.

Furthermore, if the primordial plasma is contaminated by metals from the
first objects, these might contribute to the cooling rate substantially
and thus influence the electron temperature. It is an important but
difficult task to estimate this effect.

\newpage

{\LARGE Figure Captions}
\vspace{1cm}\\
{\bf Fig. 1}\hspace{.3cm}
The induced microwave anisotropy for a texture collapsing at $t_c=7.6$
is shown in the case of a reionized universe (full line) and a not ionized
universe (dashed line) as a function of the impact time $\tau$. In the
units chosen, the decoupling time is $t_{dec}=74$, so that in this case
 $z_c/z_{dec}\sim 90$. The damping in this case is approximately a
factor of 15 and the damped signal is also widened to a width of
approximately $t_{dec}$.
\vspace{1cm}\\
{\bf Fig. 2}\hspace{.3cm}
Like Fig.~1 but for $t_c = 20$, thus $z_c/z_{dec} \sim 13$. The
damping factor is approximately 5 in this case.
\vspace{1cm}\\
{\bf Fig. 3}\hspace{.3cm}
Like Fig.~1 but for $t_c = 40$, thus $z_c/z_{dec} \sim 3$. The
damping factor is approximately 2 in this case.
\vspace{1cm}\\
{\bf Fig. 4}\hspace{.3cm}
Like Fig.~1 but for $t_c = 100> t_{dec}$. No damping occurs in
 this case. (The two curves overlay.)
\vspace{1cm}\\
{\bf Fig. 5}\hspace{.3cm}
The limiting $\rho_i/\rho_B =q$ is given in units of $10^{-7}$ which
leads to enough ($\ge n_B$) ionizing photons to reionize the universe
for a given radiation temperature $T_i$ (plotted in units of the
ionization energy $\De=13.6eV$, heavy line. This limit is compared
with the limits on background radiations at different wavelengths
given in Table~1 (crosses connected with light lines). The lower
limit applies if the ionizing radiation was emitted until today, $z=0$.
The upper limit applies if the emission of ionizing radiation seized
at $z=10$. The remaining dashed region gives the allowed range of
parameters in the universe was photoionized until $z=10$.
\vspace{1cm}\\
{\bf Fig. 6}\hspace{.3cm}
The plasma temperature $T_e$ is given as a function of the temperature
of the ionizing radiation for $q=10$ and $f_i$ a blackbody spectrum
with chemical potential for $z=200$ and a static solution of the
equations. This figure was produced for his diploma thesis by
P. Ehrismann\cite{Eh}.
\vspace{1cm}\\
{\bf Fig. 7}\hspace{.3cm}
The degree of ionization $x$ is given as a function of the temperature
of the ionizing radiation for $q=10$ and $f_i$ a blackbody spectrum
with chemical potential for $z=200$ and a static solution of the
equations. The value $x=1$ for small $T_i$ is unphysical. It is due
to the staticity assumption. This figure was produced for his diploma
thesis by P. Ehrismann \cite{Eh}.
\end{document}